\documentclass[12pt,a4paper,pctex32]{article}
\usepackage{isolatin1}
\usepackage{epsfig}
\def\lsim{\mathrel{\vcenter{\hbox{$<$}\nointerlineskip\hbox{$\sim$}}}}
\def\gsim{\mathrel{\vcenter{\hbox{$>$}\nointerlineskip\hbox{$\sim$}}}}
\newcommand{\nc}{\newcommand}
\nc{\figcap}[1]{\begin{quote}\refstepcounter{figure}
        {\bf Figure \thefigure}: {\small #1}\end{quote}}
\def\mpp{\; \raise1.0ex\hbox{${\scriptstyle +}$\kern-0.85em
      \raise-1.0ex\hbox{${\scriptstyle (-)}$}}\; }
\def\pmp{\; \raise1.0ex\hbox{${\scriptstyle -}$\kern-0.85em
      \raise-1.0ex\hbox{${\scriptstyle (+)}$}}\; }

\begin{document}
\baselineskip=20 pt

\title{A scheme with two large extra dimensions \\
    confronted with neutrino physics}
\author{J. Maalampi$^{1,2}$\thanks{jukka.maalampi@phys.jyu.fi}, \
 V. Sipil\"{a}inen$^3$\thanks{ville.sipilainen@helsinki.fi} \ and
 I. Vilja$^4$\thanks{vilja@utu.fi} \\
  $^1$\emph{Department of Physics, University of Jyv\"{a}skyl\"{a},
                                              Finland} \\
$^2$\emph{Helsinki Institute of Physics, FIN-00014 University of Helsinki,
                                              Finland} \\
$^3$\emph{Department of Physics, FIN-00014 University of Helsinki,
                                              Finland} \\
$^4$\emph{Department of Physics, FIN-20014 University of Turku,
                                              Finland}}
\date{November 25, 2002}
\maketitle

\begin{abstract}
We investigate a particle physics model in a six-dimensional
spacetime, where two extra dimensions form a torus. Particles with
Standard Model charges are confined by interactions with a scalar
field to four four-dimensional branes, two vortices accommodating
ordinary type fermions and two antivortices accommodating mirror
fermions. We investigate the phenomenological implications of
this multibrane structure by confronting the model with neutrino
physics data.
\end{abstract}

\vspace{-14cm}
\hspace{11.5cm}HIP-2002-35/TH

\thispagestyle{empty}
\newpage

\section{Introduction}

The idea that the physical spacetime might have more than three
spatial dimensions dates back to the beginning of the 20th century
and to the attempts of Nordstr\"{o}m, Kaluza and Klein to unify
gravity and electromagnetism \cite{0}. These early ideas have been
resurrected in the present-day string theories which unify gauge
theories and gravity in a ten or eleven dimensional spacetime.
Extra dimensions were originally thought to be compactified to
extremely short distances. However, according to some new ideas
developed within the string theories some of them could be
relatively large and could have phenomenological consequences testable
in physics experiments.

 It has been realized that the
presence of large extra compact spatial dimensions could make the
fundamental Planck scale $M_\ast \, $, that is the Planck
scale associated with the higher-dimensional spacetime, much smaller than it
is in the  ordinary four-dimensional world \cite{1}--\cite{3}.
This would offer an attractive solution to the hierarchy problem,
i.e.\ the disparity of the fundamental energy scales of weak
interactions and gravity. The weakness of the gravity as seen in
our visible four-dimensional spacetime, or in other words the
largeness of the effective four-dimensional Planck scale
$M_{Pl}\simeq 10^{19}$ GeV, could be understood to follow from the
spreading of the gravity force lines into the extra dimensions.
Gravitational interactions have not yet been tested at short
distances, and at present experiments still allow some of these
extra dimensions to be as large as tenths of a millimeter
\cite{4,5}.  In order to have the Planck scale $M_\ast \, $ at the
TeV range, as the solution of the hierarchy problem requires, at least
two extra dimensions are necessarily needed \cite{1}.

The basic assumption of the extra dimension scenario is that all
the fields charged under the Standard Model (SM) gauge group are
localized on a brane, the familiar 3+1-dimensional space-time,
embedded in the 4+$n$-dimensional space, called the bulk \cite{1}.
This is not only true for the SM non-singlet fermions and scalars,
but also for the SM gauge bosons, which are attached to the branes
e.g.\ due to a gauge trapping mechanism \cite{2}.
In string theory the SM particles correspond to open strings whose
ends are stuck to the branes, while gravitons and other particles
that do not carry SM charges correspond to closed strings free to
propagate in  the extra dimensions (see e.g.\ Refs.\
\cite{2,5b,5c}).

The possible unlocalizability of right-handed neutrinos ($\nu_R$)
in extra-dimension scenarios
was emphasized in \cite{6,7}. This feature may turn out to be of
crucial importance from the neutrino physics point of view. The
Yukawa coupling between the left-handed neutrino $\nu_L$, residing
on the brane, and  the right-handed neutrino $\nu_R$, residing
wherever in the spacetime, is suppressed by a factor
of the order of $M_\ast /M_{Pl}$
due to the scarce overlapping of the corresponding wave
functions. This would offer a new and elegant explanation for the
lightness of neutrinos. Perhaps the most popular conventional
explanation for the lightness of neutrinos is given by the seesaw
mechanism, which requires the existence of a new mass scale around
$10^{12}$ GeV or higher. In the extra-dimension scheme such a high
mass scale is neither needed nor naturally appears but small
neutrino masses follow from the suppression of the Yukawa
couplings due to the large bulk volume.

The possible existence of bulk neutrinos implies another novel
feature. The momenta of bulk fields are quantized in the
compactified extra dimensions, and each of these quantum states
corresponds to a massive state, a Kaluza-Klein excitation, at the
brane world. Hence a bulk neutrino would appear as an infinite
tower of massive Dirac neutrinos in our four-dimensional
spacetime. These new degrees of freedom would hugely enrich the
neutrino mass spectrum and neutrino oscillation patterns,
as discussed in many recent papers, see e.g.\
\cite{9}.

In our previous work \cite{40} we investigated a model with one large
extra dimension paying a particular attention to the effects
of the extra dimension in neutrino physics. The localization
 of the SM fermions in the four-dimensional space was achieved
 by introducing a five-dimensional scalar field with
 position dependent expectation value having a domain-wall
 profile in the extra dimension \cite{41,42}. Due to the
 periodicity of the scalar field in the compactified fifth
 dimension, there necessarily exists, in addition to the ordinary SM brane,
 another brane, a mirror brane,
in the bulk. We constructed a concrete mathematical framework
that appropriately described this two-brane scheme. We extracted
a finite neutrino mass matrix, of which the massive Kaluza-Klein
modes were integrated out, and studied neutrino masses and mixing
from phenomenological point of view.
It was found out that both the neutrino properties and the size of
the extra dimension, $R$, are in this scheme severely constrained
by experimental data. Especially, $R$ turned out to be bounded
from below due to the upper limit of the electron neutrino mass set by
the tritium beta decay experiments. The mixing between the
ordinary left-handed neutrino and the inert left-handed bulk
neutrino is fairly small if $R \lsim 10^{-3}$ mm, but maximal
mixing becomes possible for a larger $R$. A large radius, $R \gsim
10^{-2}$ mm, seemed also to be favored by cosmological arguments.
Finally, all the active-sterile neutrino mixing schemes considered
as possible solutions to the solar and atmospheric neutrino
anomalies were also found realizable in this model.

As mentioned before, the hierarchy problem cannot
be solved if there is only one extra dimension.
In this paper we therefore extend our previous analysis
to the case of two extra dimensions.
The two-dimensional extra space is assumed to be compactified on a torus.
 The torus geometry allows for a flat metric, which makes
various mathematical manipulations quite straightforward.

We are interested in figuring out in which respect this more
realistic six-dimensional scheme differs from our previously
studied five-dimensional model. In particular, we would like to
find out how the requirement of sufficiently large extra
dimensions -- so as to solve the hierarchy problem -- restricts
the parameter space of neutrino physics, that is, mixing angles
and squared mass differences.

We will use a similar field-theoretical fermion-trapping mechanism
as used in earlier studies \cite{1,43,44}. We introduce a
six-dimensional scalar field that has a vortex-like solution, the
field vanishing at the center of the vortex \cite{45}. The SM
fermions, assumed to be coupled to the scalar field, are then
trapped in the core of the four-dimensional vortex, the brane.
The periodicity of the scalar field implies that the vortex must have
two vis-\`{a}-vis antivortices in both of the extra toroidal
dimensions. Consequently, the antivortices require then
existence of two vortices, so the vortex configuration
consists of two vortices and two antivortices. In
other words, besides ``our'' SM brane one has another brane where
particles have usual chiralities and two mirror branes containing
mirror particles with chiralities opposite to those of their SM
counterparts. In addition to gravitons, only right-handed ordinary
neutrinos and their opposite-handed mirror partners are free to
propagate in the higher-dimensional bulk. Note that our way
of reasoning here is perfectly analogous to the case of one extra
dimension \cite{40}, except that now the number of branes and
mirror branes is doubled.

Note that the vortex structure on a sphere would be simpler with
just one vortex-antivortex pair. However, the curvature of
a sphere is non-zero and therefore no fermion zero modes exist
on a sphere \cite{45b}. As a consequence the bulk (inert)
neutrino structure and the interplay between the bulk
and the brane (active) neutrinos would be much more complicated.

In the following Section we shall introduce the mathematical
framework of  our six-dimensional scheme and study
its effective action in the four-dimensional spacetime.
The neutrino mass matrix is derived
and a phenomenological analysis of neutrino
physics of the model is performed in Section 3. Section 4 is
devoted to a summary and conclusions.

\section{Formulation of the model}

In this section we consider a concrete six-dimensional model that
exhibits the feature of localizing the left-handed neutrinos on
branes and the right-handed mirror neutrinos on mirror branes
while letting their gauge inert opposite-handed companions to live
in the entire higher-dimensional bulk. The localization of the
gauge active neutrinos is achieved by the field-theoretical method
that was discussed in e.g.\ Ref.\ \cite{1} and that we
applied in our previous analysis of a five-dimensional scenario
\cite{40}. The two extra dimensions are assumed to be compactified
on a torus with radii $R_1$ and $R_2$. A point in the
six-dimensional space is denoted as $(x^\mu ,y_1 ,y_2)$, where
$x^\mu$ stands for ordinary spacetime coordinates ($\mu =0,\ldots
,3$) and $y_{1(2)} \sim y_{1(2)} +2 \pi R_{1(2)}$ are periodic
coordinates on the torus.

As mentioned above, the scalar field ($\Phi$) causing the
localization is supposed to form two vortices
and two antivortices. A vortex
solution is given by $\Phi =e^{i\theta} \phi (r)$, with $\phi
(0)=0$ and $\phi (r=\mbox{large}) \simeq $ const., where $r$ is a
radial coordinate \cite{45}. For antivortices one simply changes
the sign of $\theta$. As will be seen soon, neutrinos are trapped
inside a vortex, the throat of which can be viewed as a
four-dimensional brane.

The neutrino sector of our model consists of six-dimensional
chiral spinors,  a left-handed spinor $\Psi_1$ and a right-handed
spinor $\Psi_2$, both associated with active neutrinos, together
with a spinor $N$, which is associated with an inert neutrino and
which, as will be seen later, can be chosen right-handed. The
fields described by the spinors $\Psi_1$ and $\Psi_2$ are confined
to branes or mirror branes, whereas $N$ exists everywhere in the
bulk. This choice is a minimal set-up needed to build a successful
model on a torus. However, for technical reasons we will keep for
a while both chiral components of the spinor fields in our
considerations. Contrary to the five-dimensional case \cite{40},
two active fields have now to be
introduced, in order to accomplish the localization of the fields.

The six-dimensional action of our model is taken to be
\begin{eqnarray}  \label{alku}
   S & = & \int d^4 xd^2 y[i \overline{\Psi}_1 \Gamma^A \partial_A \Psi_1
         +i \overline{\Psi}_2 \Gamma^A \partial_A \Psi_2
      +(g \Phi \overline{\Psi}_1 \Psi_2 +h.c.) \nonumber \\
     & & +i\overline{N} \Gamma^A \partial_A N
       +(\kappa H^\ast \overline{N} \Psi_1 +h.c.)] \ ,
\end{eqnarray}
where $A=0,\ldots ,5$, $\kappa$ is a
dimensionful Yukawa coupling,
and $H$ corresponds to the ordinary Higgs field. The gamma
matrices read
\begin{equation}
\Gamma^\mu =\left( \begin{array}{cc} 0 & \gamma^\mu
      \\ \gamma^\mu & 0 \end{array} \right), \ \
\Gamma^4 =\left( \begin{array}{cc} 0 & 1_4
      \\ -1_4& 0 \end{array} \right),\ \
\Gamma^5 =-i \left( \begin{array}{cc} 0 & \gamma^5
      \\ \gamma^5 & 0 \end{array} \right)\ ,
\end{equation}
where $\gamma^\mu$'s are the usual four-dimensional gamma
matrices, and $\gamma^5 =i \gamma^0 \gamma^1 \gamma^2 \gamma^3$.
It is easy to show that $\Gamma^A$'s fulfil the six-dimensional
Clifford algebra $\{\Gamma^A ,\Gamma^B \} =2g^{AB} I_8$, with
$g^{AB} =\mbox{diag}(1,-1,-1,-1,-1,-1)$. Furthermore, by defining the
six-dimensional chirality operators, $P_\pm =\frac{1}{2} (I_8 \pm
\Gamma^7)$, where $\Gamma^7 =- \Gamma^0 \Gamma^1 \cdots \Gamma^5
=\mbox{diag}(1_4 ,-1_4)$, one is able to decompose the six-dimensional
spinors as $\Psi_1 =(\Psi_{1+} \ \Psi_{1-} )^T$ and similarly for
$\Psi_2$ and $N$\footnote{See e.g.\ Ref.\ \cite{46} for an
earlier discussion on six-dimensional diracology, chiralities in
six and four dimensions, and other related issues.}. The subscript
+ denotes a right-handed and the subscript $-$ a left-handed
spinor component in the six-dimensional sense.

The reader should note that we have not presented the gauge
boson contributions, free scalar parts, nor the non-neutrino
(fermion) parts in the action (\ref{alku}), but only the parts
essential to read out the neutrino couplings. Gauge bosons are,
as mentioned earlier in this paper, trapped to the vortices,
so that, effectively, each vortex has its own SM gauge bosons,
which couple directly only to fermions on the very same brane.
Free scalar parts, on the other hand, are the standard ones,
and the non-neutrino parts are out of the scope of the present
paper.

Let us further point out some peculiar features of our action. The form of
the action (\ref{alku}) follows from the imposition of a global
U(1) symmetry, under which the fields $\Psi_1$, $\Psi_2$, $\Phi$,
$N$ and $H$ are taken to have quantum numbers 1, -1, 2, 0 and 1,
respectively. Especially, as will be seen below, the term $\Phi
\overline{\Psi}_1 \Psi_2$ takes care of the localization of the
active neutrinos on the branes. Terms like $\Phi \overline{N} N$,
$\Phi \overline{\Psi}_i \Psi_i$, $\overline{\Psi}_i \Psi_j \ (i
\neq j)$, $H^\ast \overline{N} \Psi_2$
and $H \overline{N} \Psi_1$
are forbidden by the U(1) symmetry.
Mass terms $\overline{\Psi}_i \Psi_{i}$ and
$\overline{N} N$ as well as the bulk neutrino Majorana
mass term  $\sim \overline{N} N^c $,
allowed by the symmetry, do not survive
as we will put $\Psi_{1+}=\Psi_{2-}=N_- =0$ later on.
Combinations $\Phi^{(*)} \overline{\Psi}_{1(2)} \Psi_{1(2)}^c$
, $\overline{\Psi}_i \Psi_{j}^c \ (i \neq j)$
and $H \overline{N} \Psi_2$,
on the other hand, vanish when the SU(2) structure
is taken into account: $\Psi_i$'s and $H$ are
the neutral parts of SU(2) doublets whereas
$N$ is a SU(2) singlet.

We start to analyze our model by studying the localization of the
brane fermions.
The equations of motion of the fields $\Psi_{1,2}$ are
\begin{eqnarray}  \label{c1}
    i\Gamma^A \partial_A \Psi_1 +g \Phi \Psi_2 & = & 0 \ ,
  \nonumber \\
    i\Gamma^A \partial_A \Psi_2 +g^* \Phi^* \Psi_1 & = & 0 \ ,
\end{eqnarray}
where a term $\kappa^* HN$ has been
neglected, since it may be regarded as a small perturbation
in the lowest order approximation.
Using polar coordinates ($y_1 =r \cos \theta,y_2 =r \sin \theta
$), one has
\begin{equation}
  \Gamma^4 \partial_4 +\Gamma^5 \partial_5 =
  \left( \begin{array}{cc}
   0 & e^{-i\gamma^5 \theta} \partial_r
   -\frac{i}{r} \gamma^5 e^{-i\gamma^5 \theta} \partial_\theta \\
     -e^{i\gamma^5 \theta} \partial_r
   -\frac{i}{r} \gamma^5 e^{i\gamma^5 \theta} \partial_\theta & 0
   \end{array} \right)\ ,
\end{equation}
and thus Eqs.\ (\ref{c1}) can be written in terms of the chiral
components of $\Psi_{1,2}$ as
\begin{eqnarray}  \label{c2}
  -i\gamma^\mu \partial_\mu \Psi_{1-} & = &
   i e^{-i\gamma^5 \theta} \partial_r \Psi_{1-}
   +\frac{1}{r} \gamma^5 e^{-i\gamma^5 \theta} \partial_\theta
   \Psi_{1-} +g e^{i\theta} \phi (r) \Psi_{2+} \ , \nonumber \\
   -i\gamma^\mu \partial_\mu \Psi_{1+} & = &
   -i e^{i\gamma^5 \theta} \partial_r \Psi_{1+}
   +\frac{1}{r} \gamma^5 e^{i\gamma^5 \theta} \partial_\theta
   \Psi_{1+} +g e^{i\theta} \phi (r) \Psi_{2-} \ , \nonumber \\
   -i\gamma^\mu \partial_\mu \Psi_{2-} & = &
   i e^{-i\gamma^5 \theta} \partial_r \Psi_{2-}
   +\frac{1}{r} \gamma^5 e^{-i\gamma^5 \theta} \partial_\theta
   \Psi_{2-} +g^* e^{-i\theta} \phi (r) \Psi_{1+} \ , \\
   -i\gamma^\mu \partial_\mu \Psi_{2+} & = &
   -i e^{i\gamma^5 \theta} \partial_r \Psi_{2+}
   +\frac{1}{r} \gamma^5 e^{i\gamma^5 \theta} \partial_\theta
   \Psi_{2+} +g^* e^{-i\theta} \phi (r) \Psi_{1-} \ . \nonumber
\end{eqnarray}
Here we consider the localization on a vortex (i.e.\ $\Phi
=e^{i\theta} \phi (r)$); the case of an antivortex localization
proceeds along the same lines, with just the chiralities appropriately
reversed (see below).

At this point we return to our original chiral picture by setting
$\Psi_{1+}=\Psi_{2-}=0$.  Looking at the first and fourth
of equations of (\ref{c2})\footnote{The second and third equation would not
have entered at all if we had used chiral fields from the beginning.},
it is now natural to require that $\gamma^5 \Psi_{1-}=-\Psi_{1-}$
and $\gamma^5 \Psi_{2+}=-\Psi_{2+}$. This means that
$\Psi_{1-}=\frac{1}{2}(1-\gamma^5)\Psi_{1-}\equiv \Psi_{1--}$, the
second sign in the subscript
referring to the ordinary four-dimensional chirality of
$\Psi_1$ (i.e., $\Psi_1$ is left-handed in four-dimensional
sense). Similarly $\Psi_{2+}=\Psi_{2+-}$, i.e., the
four-dimensional chirality of $\Psi_2$ is left-handed, while its
six-dimensional chirality is ``right-handed''.
Furthermore, a solution of the equations of motion apparently
exists if $\partial_\theta \Psi_{1-}=
\partial_\theta \Psi_{2+}=0$.
Eqs.\ (\ref{c2})
now reduce to
\begin{eqnarray}
    -i\gamma^\mu \partial_\mu \Psi_{1--} & = &
   e^{i\theta}(i \partial_r \Psi_{1--}
   +g \phi (r) \Psi_{2+-}) \ , \nonumber \\
    -i\gamma^\mu \partial_\mu \Psi_{2+-} & = &
   e^{-i\theta}(-i \partial_r \Psi_{2+-}
   +g^* \phi (r) \Psi_{1--}) \ .
\end{eqnarray}
Writing $\Psi_{1--}=\varphi_{1--}(x)f_1 (r)$ and
$\Psi_{2+-}=\varphi_{2+-}(x) f_2 (r)$, where $f_{1,2}$
are real functions, and $\varphi_{1--}=v\varphi_{2+-}$,
one ends up with the equations
\begin{eqnarray}  \label{c3}
    -f_1 (r)(i\gamma^\mu \partial_\mu \varphi_{1--}) & = &
   e^{i\theta} \varphi_{1--}(i \partial_r f_1 (r)
   +g v^{-1} \phi (r) f_2 (r)) \ , \nonumber \\
   -v^{-1} f_2 (r)(i\gamma^\mu \partial_\mu \varphi_{1--}) & = &
   e^{-i\theta} \varphi_{1--}(-i v^{-1} \partial_r f_2 (r)
   +g^* \phi (r) f_1 (r)) \ .
\end{eqnarray}
The l.h.s.\ of these equations vanish due to the
usual four-dimensional Dirac equation for a
massless field. If we
take $v=\pm i e^{i\arg (g)}$, the r.h.s.\ can be easily
shown to have solutions behaving like\footnote{This requires that
$f_1 (0)=f_2 (0)$.} $f_{1,2}(r) \sim \exp (\pm |g| \int_{0}^r ds
\phi(s))$, where, in order to confine $f_{1,2}(r)$ to the vicinity
of $r=0$, the plus (minus) sign must be chosen for
$\phi(r)<0 \ (>0)$. This concludes our demonstration of the trapping
of the spinor fields $\Psi_1$ and $\Psi_2$ on a vortex by the
scalar field $\Phi$.

Two comments are worthwhile at this stage. First, it is
straightforward to see what will happen in the antivortex case.
Substituting $\Phi =e^{-i\theta} \phi (r)$ to Eqs.\ (\ref{c1}) and
(\ref{c2}), one sees immediately that now a natural choice is
$\gamma^5 \Psi_{1-(2+)} =+\Psi_{1-(2+)}$, and it is hence the
right-handed components $\Psi_{1-+}$ and $\Psi_{2++}$ that get
localized on an antivortex. This is of course what one would have
intuitively expected to happen a priori. Second, as can be
deduced from the second and third of equations of (\ref{c2}), the
condition $\Psi_{1+}=\Psi_{2-}=0$ not only makes life easier, but
is necessary if \emph{only} the left-handed neutrinos (ordinary
active neutrinos) are wanted to get trapped on a vortex and
right-handed neutrinos (active mirror neutrinos) on an antivortex.
As is known, there is so far no indication of the existence of mirror
neutrinos in our four-dimensional world \cite{47}.

One may now write somewhat symbolically
\begin{eqnarray}   \label{c4}
  \Psi_1 & = & \sqrt{\delta (\bar{y})}
   \left( \begin{array}{c} 0 \\ \varphi_{1--}(x)
   \end{array} \right)
  +\sqrt{\delta (\bar{y}-\bar{y}_3)}
   \left( \begin{array}{c} 0 \\ \varphi_{3--}(x)
   \end{array} \right)
  +\sqrt{\delta (\bar{y}-\bar{y}_2)}
   \left( \begin{array}{c} 0 \\ \varphi_{2-+}(x)
   \end{array} \right) \nonumber \\
  & & +\sqrt{\delta (\bar{y}-\bar{y}_4)}
   \left( \begin{array}{c} 0 \\ \varphi_{4-+}(x)
   \end{array} \right) \ , \nonumber \\
  \Psi_2 & = & \sqrt{\delta (\bar{y})}
   \left( \begin{array}{c} \varphi_{1--}(x) \\ 0
   \end{array} \right)
   +\sqrt{\delta (\bar{y}-\bar{y}_3)}
   \left( \begin{array}{c} \varphi_{3--}(x) \\ 0
   \end{array} \right)
   +\sqrt{\delta (\bar{y}-\bar{y}_2)}
   \left( \begin{array}{c} \varphi_{2-+}(x) \\ 0
   \end{array} \right) \nonumber \\
  & & +\sqrt{\delta (\bar{y}-\bar{y}_4)}
   \left( \begin{array}{c} \varphi_{4-+}(x) \\ 0
   \end{array} \right) \ ,
\end{eqnarray}
where the branes (vortices) are situated at $\bar{y}_1(=0)$ and
$\bar{y}_3$, and the mirror branes (antivortices) at $\bar{y}_2$
and $\bar{y}_4$. The form of $\Psi_2$ is due to the requirement
that (see above) $\Psi_{1-}(r=0)=\Psi_{2+}(r=0)$ modulo a phase
factor that can be absorbed to $\Phi$ (or alternatively to $g$)
in the original action of Eq.\ (\ref{alku}). The issue of the
Higgs field is now more involved than in the five-dimensional case
\cite{40}, and the global behavior of the field is not known in
general. However, since $H$ can be locally shown to get trapped on
a vortex or an antivortex \cite{1}, we write
\begin{equation}  \label{c5}
   H=\sqrt{\delta (\bar{y})}h_1 (x) +
     \sqrt{\delta (\bar{y}-\bar{y}_2)}h_2 (x) +
     \sqrt{\delta (\bar{y}-\bar{y}_3)}h_3 (x) +
     \sqrt{\delta (\bar{y}-\bar{y}_4)}h_4 (x) \ ,
\end{equation}
where $h_i$ can be viewed as the vev of the Higgs field in the $i$th
brane. By substituting Eqs.\ (\ref{c4}) and (\ref{c5}), together
with a Kaluza-Klein expansion
\begin{equation}
   \left( \begin{array}{c} N_+  \\
     N_- \end{array} \right)
   = \frac{1}{2 \pi \sqrt{R_1 R_2}} \sum_{k,l=-\infty}^{\infty}
   \left( \begin{array}{c} N_{+}^{kl} (x) \\
     N_{-}^{kl} (x) \end{array} \right)
    e^{iky_1 /R_1 +ily_2 /R_2 } \ ,
\end{equation}
to Eq.\ (\ref{alku}), one obtains
\begin{eqnarray}   \label{vali}
  S & = & \int d^4 x \left\{ i2\overline{\varphi}_{1--}
      \gamma^\mu \partial_\mu \varphi_{1--} +
     i2\overline{\varphi}_{2-+}
      \gamma^\mu \partial_\mu \varphi_{2-+} +
     i2\overline{\varphi}_{3--}
      \gamma^\mu \partial_\mu \varphi_{3--} \right. \nonumber \\
    & & +i2\overline{\varphi}_{4-+}
      \gamma^\mu \partial_\mu \varphi_{4-+}
      +\sum_{k,l=-\infty}^{\infty} \left[
    i\overline{N}_{+}^{kl} \gamma^\mu \partial_\mu N_{+}^{kl}
    \right. \nonumber \\   & &
  +\overline{N}_{++}^{kl}\left(\frac{k}{R_1}-\frac{il}{R_2}
   \right)N_{+-}^{kl}
   +\overline{N}_{+-}^{kl}\left(\frac{k}{R_1}+\frac{il}{R_2}
   \right)N_{++}^{kl} \nonumber \\
   & &  + u(h_{1}^* \overline{N}_{++}^{kl}
    \varphi_{1--} + h_{2}^* z^{kl}_2 \overline{N}_{+-}^{kl}
    \varphi_{2-+}  \\
   & & \left. \left.
   +h_{3}^* z^{kl}_3 \overline{N}_{++}^{kl} \varphi_{3--}
   +h_{4}^* z^{kl}_4 \overline{N}_{+-}^{kl} \varphi_{4-+}
   +h.c.) \right] \right\} \nonumber \ ,
\end{eqnarray}
where
\begin{equation}
u=\frac{\kappa}{2 \pi \sqrt{R_1 R_2}} \ ,
\end{equation}
$\kappa$ is taken to be real, and $z^{kl}_j = e^{-ik
\vartheta_j -il \omega_j}$, $(\vartheta_j ,\omega_j)$ indicating
the position of the $j$th brane in the extra dimensions (with
$\vartheta_1 =\omega_1 =0$). Our analysis complies with a
``minimalistic'' view in the sense that we have set $N_- =0$,
i.e., $N$ is taken to be a right-handed chiral field. We remind that
this choice made the Dirac and Majorana bulk
neutrino mass terms to disappear in the original action (\ref{alku}).

In order to integrate out the massive Kaluza-Klein modes, we
proceed along the same lines as we followed in our previous study
\cite{40}. This time, however, the ensuing equations are quite
lengthy and intransparent and they are mostly
omitted in the following presentation. The
first ten $k,l$-dependent terms of the action (\ref{vali})
disappear due to the equations of motion of the Kaluza-Klein
excitations $N^{kl}_+$, for example,
\begin{equation}
  N_{++}^{kl}=-\left(\frac{k}{R_1}+\frac{il}{R_2}
   \right)^{-1} (i\gamma^\mu \partial_\mu N_{+-}^{kl} +
   u h_{2}^* z^{kl}_2 \varphi_{2-+} +u h_{4}^* z^{kl}_4\varphi_{4-+})
   \ ,
\end{equation}
and similarly for $N_{+-}^{kl}$
(here obviously $(k,l) \neq (0,0)$). Substituting these twice
to the remaining four h.c.-terms of the action, neglecting
higher order derivatives of the bulk fields, and regarding
$h_i$'s as constants, one obtains
\begin{eqnarray}        \label{vali2}
  \mathcal{L}_{(k,l) \neq (0,0)} & = &
   -u^2 R_1 (\Xi_{\vartheta}(\varphi_i) + h.c.)
      +iu^2 \rho R_1 (\Xi_{\omega}(\varphi_i) - h.c.) \nonumber \\
   & & +iu^2 R_{1}^2 [|h_1|^2 s(0,0,\rho) \overline{\varphi}_{1--}
         \gamma^\mu \partial_\mu \varphi_{1--} +
          |h_2|^2 s(0,0,\rho) \overline{\varphi}_{2-+}
         \gamma^\mu \partial_\mu \varphi_{2-+}
          \nonumber  \\
    & & +|h_3|^2 s(0,0,\rho) \overline{\varphi}_{3--}
         \gamma^\mu \partial_\mu \varphi_{3--} +
          |h_4|^2 s(0,0,\rho) \overline{\varphi}_{4-+}
         \gamma^\mu \partial_\mu \varphi_{4-+}
          \nonumber  \\
    & & +(h_1 h_{3}^* s(\vartheta_{31},\omega_{31},\rho)
         \overline{\varphi}_{1--}
         \gamma^\mu \partial_\mu \varphi_{3--}  \\
    & &  +h_2 h_{4}^* s(\vartheta_{42},\omega_{42},\rho)
         \overline{\varphi}_{2-+}
         \gamma^\mu \partial_\mu \varphi_{4-+} +h.c.)] \ , \nonumber
\end{eqnarray}
where $\vartheta_{ij}=\vartheta_i -\vartheta_j$,
$\omega_{ij}=\omega_i -\omega_j$,
\begin{eqnarray}   \label{apu1}
    s_{\vartheta (\omega)} (\vartheta ,\omega ,\rho) & = &
     \sum_{(k,l) \neq (0,0)} \frac{k(l)}{k^2 +\rho^2 l^2}
    e^{-ik \vartheta -il \omega} \ , \nonumber \\
     s(\vartheta ,\omega ,\rho) & = &
     \sum_{(k,l) \neq (0,0)} \frac{1}{k^2 +\rho^2 l^2}
    e^{-ik \vartheta -il \omega} \ , \nonumber \\
   \Xi_{\omega(\vartheta)}(\varphi_i) & = &
  \pmp h_2 h_{1}^* s_{\vartheta(\omega)} (\vartheta_{12},\omega_{12},
      \rho) \overline{\varphi}_{2-+} \varphi_{1--} \nonumber \\
   & & \pmp h_4 h_{1}^* s_{\vartheta(\omega)} (\vartheta_{14},\omega_{14},
      \rho) \overline{\varphi}_{4-+} \varphi_{1--}
       \nonumber \\
   & & +h_3 h_{2}^* s_{\vartheta(\omega)} (\vartheta_{23},\omega_{23},
      \rho) \overline{\varphi}_{3--} \varphi_{2-+}  \\
   & & \pmp h_4 h_{3}^* s_{\vartheta(\omega)} (\vartheta_{34},\omega_{34},
      \rho) \overline{\varphi}_{4-+} \varphi_{3--} \ ,  \nonumber
\end{eqnarray}
and $\rho=R_1 /R_2$. Combining then Eqs.\ (\ref{vali2})
and (\ref{vali}) (with $k=l=0$), we end up with the
effective Lagrangian
\begin{eqnarray}   \label{vali3}
    \mathcal{L}_{eff} & = &
    i\overline{\varphi}_1 \gamma^\mu \partial_\mu \varphi_1
    + i\overline{\varphi}_2 \gamma^\mu \partial_\mu \varphi_2
    + i\overline{\varphi}_3 \gamma^\mu \partial_\mu \varphi_3
    + i\overline{\varphi}_4 \gamma^\mu \partial_\mu \varphi_4
    + i\overline{N}_{+} \gamma^\mu \partial_\mu N_{+} \nonumber \\
  & & + u(h_{1}^* n_1 \overline{N}_{++} \varphi_1
    + h_{2}^* n_2 \overline{N}_{+-} \varphi_2
    + h_{3}^* n_3 \overline{N}_{++} \varphi_3
    + h_{4}^* n_4 \overline{N}_{+-} \varphi_4 +h.c.) \nonumber \\
  & &   -u^2 R_1 (\Xi_{\vartheta}(n_i \varphi_i) + h.c.)
      +iu^2 \rho R_1 (\Xi_{\omega}(n_i \varphi_i) - h.c.)  \\
  & & +i(\tilde{a} \overline{\varphi}_1 \gamma^\mu \partial_\mu \varphi_3
    + \tilde{b} \overline{\varphi}_2 \gamma^\mu \partial_\mu \varphi_4
      + h.c.) \ , \nonumber
\end{eqnarray}
where we have suppressed the chirality indices of $\varphi_i$'s
and the Kaluza-Klein indices of $N$'s, rescaled $\varphi_i$ fields
with $n_i =(2+u^2 R_1^2 |h_i|^2 s(0,0,\rho))^{-1/2}$, and we have denoted
\begin{eqnarray}
  \tilde{a} & = & u^2 R_{1}^2
   h_1 h_{3}^* n_1 n_3 s(\vartheta_{31},\omega_{31},\rho) \ , \nonumber \\
  \tilde{b} & = & u^2 R_{1}^2
   h_2 h_{4}^* n_2 n_4 s(\vartheta_{42},\omega_{42},\rho) \ .
\end{eqnarray}
Note that $s(0,0,\rho)$ is formally a divergent function.
Therefore physical cut-offs $k < R_1 M_*$, $l < R_2 M_*$ have been
introduced in the sums. This leads to the expression
\begin{equation}
\rho s(0,0,\rho) \simeq 2\pi \ln {M_{Pl}\over M_*}
+ \pi + \frac{\pi^2}{3} \left(\rho + \frac 1\rho \right)
- \frac 2M_* \left(\frac 1R_1 + \frac 1 R_2 \right) \ ,
\end{equation}
where $M_* = (M_{Pl}^2 /R_1R_2)^{1/4}$.

Finally, by defining
\begin{equation}   \label{apu10}
   \left( \begin{array}{c}
     \varphi_1 \\ \varphi_2 \\ \varphi_3 \\ \varphi_4
   \end{array} \right) =\frac{1}{\sqrt{2}} \left(
    \begin{array}{cccc}
   e^{i\zeta_a} & 0 & e^{i\zeta_a} & 0   \\
   0 &  e^{i\zeta_b} & 0 &  e^{i\zeta_b}  \\
   -e^{-i\zeta_a} & 0 & e^{-i\zeta_a} & 0 \\
   0 & -e^{-i\zeta_b} & 0 &  e^{-i\zeta_b} \end{array}
    \right)
     \left( \begin{array}{c}
     \xi_1 \\ \xi_2 \\ \xi_3 \\ \xi_4
   \end{array} \right) \ , \ \zeta_a =\frac{1}{2} \arg \tilde{a}
   \ , \ \zeta_b =\frac{1}{2} \arg \tilde{b} \ ,
\end{equation}
in order to get rid of the kinetic cross terms, and rescaling
$\xi_i$'s with $\tilde{n}_{1(3)}=(1\pmp |\tilde{a}|)^{-1/2}$ and
$\tilde{n}_{2(4)}=(1 \pmp |\tilde{b}|)^{-1/2}$, Eq.\ (\ref{vali3})
may be cast into the form
\begin{eqnarray} \label{Leff}
   \mathcal{L}_{eff} & = &
    i\overline{\xi}_1 \gamma^\mu \partial_\mu \xi_1
    + i\overline{\xi}_2 \gamma^\mu \partial_\mu \xi_2
    + i\overline{\xi}_3 \gamma^\mu \partial_\mu \xi_3
    + i\overline{\xi}_4 \gamma^\mu \partial_\mu \xi_4
    + i\overline{N}_{+} \gamma^\mu \partial_\mu N_{+} \nonumber \\
    & & -(\mathcal{A} \overline{\xi}_2 \xi_1
    + \mathcal{B} \overline{\xi}_4 \xi_1
    + \mathcal{C} \overline{\xi}_2 \xi_3
    + \mathcal{D} \overline{\xi}_4 \xi_3 + h.c.) \\
    & & -(\mathcal{E} \overline{N}_{++} \xi_1
    + \mathcal{F} \overline{N}_{+-} \xi_2
    + \mathcal{G} \overline{N}_{++} \xi_3
    + \mathcal{H} \overline{N}_{+-} \xi_4 + h.c.) \ , \nonumber
\end{eqnarray}
where
\begin{eqnarray}
   \mathcal{E} & = & -\frac{u \tilde{n}_1}{\sqrt{2}}
     (h_{1}^* n_1 e^{i\zeta_a}-h_{3}^* n_3 e^{-i\zeta_a})
    \ , \nonumber \\
    \mathcal{F} & = & -\frac{u \tilde{n}_2}{\sqrt{2}}
     (h_{2}^* n_2 e^{i\zeta_b}-h_{4}^* n_4 e^{-i\zeta_b})
    \ , \nonumber \\
    \mathcal{G} & = & -\frac{u \tilde{n}_3}{\sqrt{2}}
     (h_{1}^* n_1 e^{i\zeta_a}+h_{3}^* n_3 e^{-i\zeta_a})
    \ ,    \\
    \mathcal{H} & = & -\frac{u \tilde{n}_4}{\sqrt{2}}
     (h_{2}^* n_2 e^{i\zeta_b}+h_{4}^* n_4 e^{-i\zeta_b})
    \ . \nonumber
\end{eqnarray}
The explicit expressions of the quantities $\mathcal{A},
\mathcal{B},\mathcal{C}$ and $\mathcal{D}$ are irrelevant for what
follows. One should notice that if the phases of $h_i$'s are
absorbed to the $\varphi_i$ fields in Eq.\ (\ref{vali}), one has,
$s(\vartheta ,\omega ,\rho)$ being real, $\zeta_a =\zeta_b =0$,
and thus also $\mathcal{E},\mathcal{F},\mathcal{G}$ and
$\mathcal{H}$ can all be made real.

\section{Neutrino phenomenology of the model }

We now move to consider the phenomenological aspects of our
scheme, in particular the effects of the multibrane structure on
neutrino physics. Apart from gravitons, only sterile neutrinos can
mediate contacts between branes and mirror branes through the bulk,
thereby affecting neutrino phenomenology at our home brane.

From the effective Lagrangian $\mathcal{L}_{eff}$ in Eq.\
(\ref{Leff}) one infers the following mass Lagrangian for
neutrinos:
\begin{equation}
   \mathcal{L}_M =-(\overline{\xi}_2 \, \overline{\xi}_4 \,
       \overline{N}_{++}) M \left(
      \begin{array}{c} \xi_1 \\ \xi_3 \\ N_{+-} \end{array}
      \right) +h.c. \ ,
\end{equation}
where the matrix $M$ is given by
\begin{equation}
    M=\left( \begin{array}{ccc}
       \mathcal{A} & \mathcal{C} & \mathcal{F} \\
       \mathcal{B} & \mathcal{D} & \mathcal{H} \\
       \mathcal{E} & \mathcal{G} & 0
     \end{array} \right) \ .
\end{equation}
Even though $M$ can in principle be diagonalized in its general
form, this task becomes more straightforward if the branes are
assumed to be distributed symmetrically on the torus\footnote{For
example $\vartheta_3 =\omega_3 =\vartheta_2 =\omega_4 =\pi$,
$\omega_2 =\vartheta_4 =0$. Then
$s(\vartheta_{31},\omega_{31},\rho)=s(\vartheta_{42},\omega_{42},\rho)=
s(\pi,\pi,\rho)$, which is easily calculable numerically.}.
We have made this assumption in the following.
Then, as can be
deduced from Eqs.\ (\ref{apu1}), $\Xi_\vartheta =\Xi_\omega =0$
and consequently $\mathcal{A}= \mathcal{B}= \mathcal{C}=
\mathcal{D}=0$. As will be seen later, even this simplified
scenario leads to diverse neutrino phenomenology. Defining
\begin{equation}   \label{apu11}
   O_\alpha = \left( \begin{array}{ccc}
    \cos \alpha & \sin \alpha & 0 \\
    -\sin \alpha & \cos \alpha & 0 \\
      0 & 0 & 1  \end{array} \right) \ , \ \
   O_\beta = \left( \begin{array}{ccc}
    \cos \beta & 0 & \sin \beta  \\
    -\sin \beta & 0 & \cos \beta  \\
      0 & 1 & 0  \end{array} \right) \ ,
\end{equation}
with
\begin{equation}
\cos \alpha =\frac{\mathcal{G}}{\sqrt{\mathcal{E}^2 +\mathcal{G}^2}}
\ , \ \
\cos \beta =\frac{\mathcal{H}}{\sqrt{\mathcal{F}^2 +\mathcal{H}^2}}
\ ,
\end{equation}
we obtain
\begin{equation}
   O_{\beta}^T M O_\alpha = \mbox{diag}(m_1,m_2,m_3) \ ,
\end{equation}
where the eigenvalues are $m_1 =0$, $m_2=\sqrt{\mathcal{E}^2
+\mathcal{G}^2}$ and $m_3=\sqrt{\mathcal{F}^2 +\mathcal{H}^2}$.
The corresponding mass eigenstates are
\begin{equation}   \label{apu12}
    \left(\begin{array}{c} \chi_1 \\ \chi_2 \\ \chi_3 \end{array}
      \right) = O_{\alpha}^T
    \left(\begin{array}{c} \xi_1 \\ \xi_3 \\ N_{+-} \end{array}
      \right) + O_{\beta}^T
    \left(\begin{array}{c} \xi_2 \\ \xi_4 \\ N_{++} \end{array}
      \right) \ .
\end{equation}
The neutrino sector thus consists of one massless and two massive
Dirac neutrinos. We emphasize that the existence of one massless
eigenstate is not an intrinsic feature of our scenario but
results from our choice of symmetrically located
branes. Another characteristic feature
of the model is the imposed
right-handed chirality of the bulk field $N$.
As mentioned before, this choice
automatically wipes out the six-dimensional Dirac
and Majorana bulk neutrino mass terms, thus removing
a usual need to rely on ad hoc arguments.

Let us remind the reader that in Eq.\ (\ref{apu12})
$\xi_1$ and $\xi_3$ are related to left-handed
active neutrinos (through e.g.\ Eq.\ (\ref{apu10})),
$\xi_2$ and $\xi_4$ to right-handed active mirror neutrinos
and the $N$ fields represent sterile bulk neutrinos
not confined to any brane.
Using Eqs.\ (\ref{apu10}), (\ref{apu11}) and (\ref{apu12}),
taking into account the field rescalings, and normalizing
correctly, the original left-handed
neutrino of ``our'' brane is then given by
\begin{equation}  \label{zzzz}
   \varphi_1 = \cos \theta \chi_{1L} + \sin \theta \chi_{2L}\ ,
\end{equation}
where
\begin{equation} \label{tassu}
   \cos \theta =\frac{\tilde{n}_1 \cos \alpha -
   \tilde{n}_3 \sin \alpha}{\sqrt{\tilde{n}_{1}^2
    + \tilde{n}_{3}^2}} \ .
\end{equation}
As $\varphi_1$ is the only active neutrino living on our brane,
only the mixing angle $\theta$ can be experimentally probed. The
angle $\beta$ is related to the right-handed fields $\varphi_2$
and $\varphi_4$, and is therefore a measurable quantity only on
the mirror branes.

It should be pointed out that the physical setting
of the neutrino mixing differs in one noteworthy
respect from the results of our previous
five-dimensional study. In \cite{40} the
ordinary left-handed neutrino was found to be
mixed with an inert left-handed bulk neutrino.
Now, as can be seen easily by repeating the
calculations similar to those leading to
Eq.\ (\ref{zzzz}), the other linear combination
(not necessarily orthogonal to Eq.\ (\ref{zzzz}))
of $\chi_{1L}$ and $\chi_{2L}$ is $\varphi_3$.
$\varphi_1$ is hence mixed with a left-handed
active neutrino living on another brane.

Similarly, as seen from Eq.\ (\ref{zzzz}), the massless
neutrino $\chi_1$ and the massive state $\chi_2$ appear as mixed
in weak interactions. There exist several empirical constraints on
their mixing angle $\theta$ and the mass $m_2$, coming from
laboratory experiments, astrophysical observations and
cosmological considerations. In Fig.\ 1 we compile (as a rough
step-function approximation sufficient for our purposes) the
existing laboratory bounds on the mixing angle $\theta$ between
the electron neutrino and a heavier neutrino in various mass
ranges of the mass $m_2$, see Refs.\ \cite{61}.
These bounds follow mainly from the lack of extra
kinematical thresholds in the spectra of various
particle decays (e.g.\ beta decay or $\pi^+
\rightarrow e^+ \nu_e$) or the nonobservation of the
heavier neutrino decays like $\nu \rightarrow \nu_e +e^- +e^+$.
One should take into
account also the upper limit of the electron neutrino mass from
the tritium beta decay experiment \cite{72},
\begin{equation}
m_{\nu_{e}} \lsim 2.2\ {\rm eV} .
\end{equation}
This bound should be fulfilled by the mass eigenvalue $m_2$,
whenever the electron neutrino is predominantly the heavier mass
eigenstate $\chi_2$. An important constraint at low $m_2$ values
is given by the oscillation limit from the Bugey disappearance
experiment \cite{72b}. The upper bound for $\sin^2 2\theta$ varies
in the range 0.02 to 0.3 for $m_2^2$ in the range 0.01 eV$^2$ to 100
eV$^2$.

\begin{figure}[ht]
\leavevmode \centering \vspace*{90mm}
\begin{picture}(0,0)(0,490)
\includegraphics{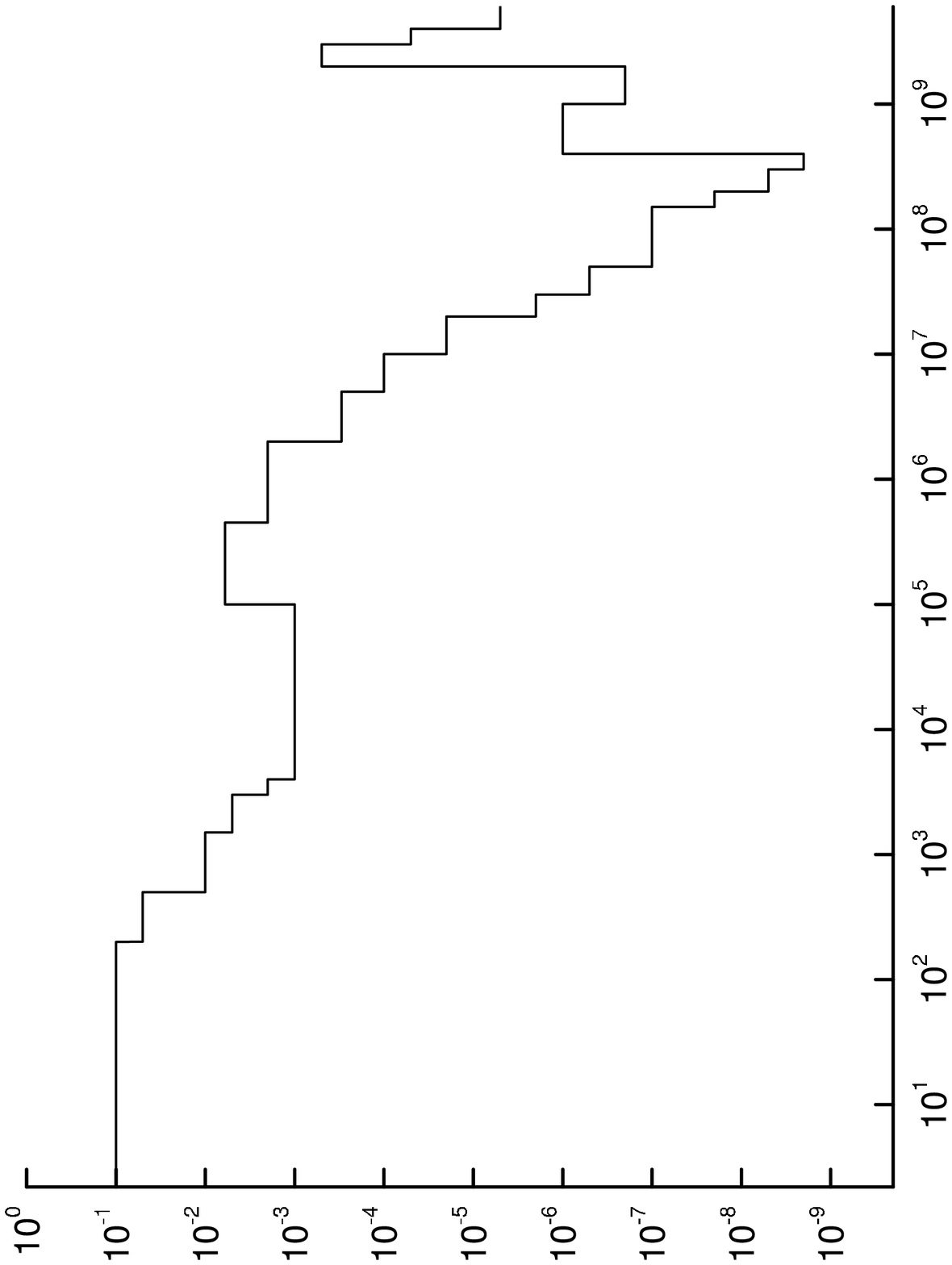}
\put(0,235){$\scriptstyle {m_2}\ {\rm (eV)}$}
\put(-210,365){$\scriptstyle \sin^2 \theta $}
\put(-50,350){Allowed region}
\end{picture}
\caption{Laboratory bounds on the mixing angle between the electron
neutrino and a sterile neutrino from laboratory measurements as a
function of the mass of the heavier mass eigenstate (the lighter
eigenstate is massless). } \label{kuva0}
\end{figure}

The active-sterile neutrino mixing is constrained also by cosmology. If the
mixing is too large, neutrino oscillations, acting as an effective
interaction, would bring sterile neutrinos into equilibrium with
the light SM particles before neutrino decoupling. The resulting
excess in energy density would endanger the standard scheme of the
nucleosynthesis of light elements \cite{73}. This leads to
the following bound for $\nu_e \leftrightarrow \nu_s$ mixing
\cite{74}:
\begin{equation}
\begin{array}{ll}  \label{shi}
    |\delta m^2| \sin^2 2\theta <5 \times 10^{-8}\
    \mbox{eV}^2 & \mbox{for}\ |\delta m^2|<4\ \mbox{eV}^2\ ,
      \\
    \sin^2 2\theta <10^{-8}\
    & \mbox{for}\ |\delta m^2|>4\ \mbox{eV}^2\ .
\end{array}
\end{equation}
(This bound is avoided if there was a suitable net lepton number
in the early universe \cite{75}.)

We will now confront our scheme with these
constraints. We will search the allowed regions of the parameter
space of the model numerically using a Monte Carlo analysis. We
vary three unknown parameters: the radii of the two extra
dimensions $R_1$ and $R_2$, and the Higgs vacuum expectation value
$h_3$. The extra dimension radii are varied from the Planck scale
to a millimeter scale and the Higgs vev between $10^{-2}$ MeV and
$10^{12}$ MeV. Both $R_i$'s and $h_3$ are randomized so that their
logarithms are evenly distributed. For $h_1$ we use the value 174
GeV, and, as can be seen from Secs.\ 2 and 3,
$h_2$ and $h_4$ do not affect neutrino physics on our brane.

In addition to these parameters the action depends on the
dimensionful Yukawa coupling $\kappa$. The natural scale for it
can be thought to be set by the higher-dimensional Planck scale
$M_*$, and therefore we write (with $M_{Pl}^2 =M_{\ast}^{4} R_1
R_2$)
\begin{equation}
\kappa=\frac{\kappa '}{M_*}=
 \kappa^\prime \left({R_1 R_2 \over M_{Pl}^2}\right)^{1/4},
\end{equation}
where $\kappa^\prime$ is a dimensionless coupling constant. The
value of $\kappa^\prime$ is not really known, but a plausible
choice would be a number relatively close to unity. We have
performed our analysis for the values $\kappa^\prime = 1$ and
$\kappa^\prime = 0.001$.

In Figs.\ 2 we present scatter plots in the $(\bar R,m_2)$-plane
(with $\bar R = \sqrt {R_1R_2}$) for the two values of the Yukawa
coupling ($\kappa'=0.001$ and 1). These figures consist solely of
``raw data'', i.e.\ neither the cosmological constraint,
Eq.\ (\ref{shi}), the limits of Fig.\ 1,
nor the Bugey constraint are taken into account.
The allowed parameter space is seen to be quite
severely constrained by the mathematical structure of the model.
Nevertheless, the characteristic extra-dimensional size may vary
from the millimeter scale some 30 orders of magnitude all the way
down to the Planck scale, and the higher mass eigenvalue $m_2$
ranges from $10^{-9}$ eV to $10^{10}$ MeV, depending on the
coupling $\kappa '$. The general trend is that small (large)
extra-dimensional size corresponds to large (small) mass. Note
that the value of $\kappa '$ affects the vertical position of the
two diagonal borderlines, whereas the wedge-like region below them
remains immutable.

\begin{figure}[ht]
\leavevmode
\centering
\vspace*{100mm}
\begin{picture}(0,0)(0,490)
\includegraphics{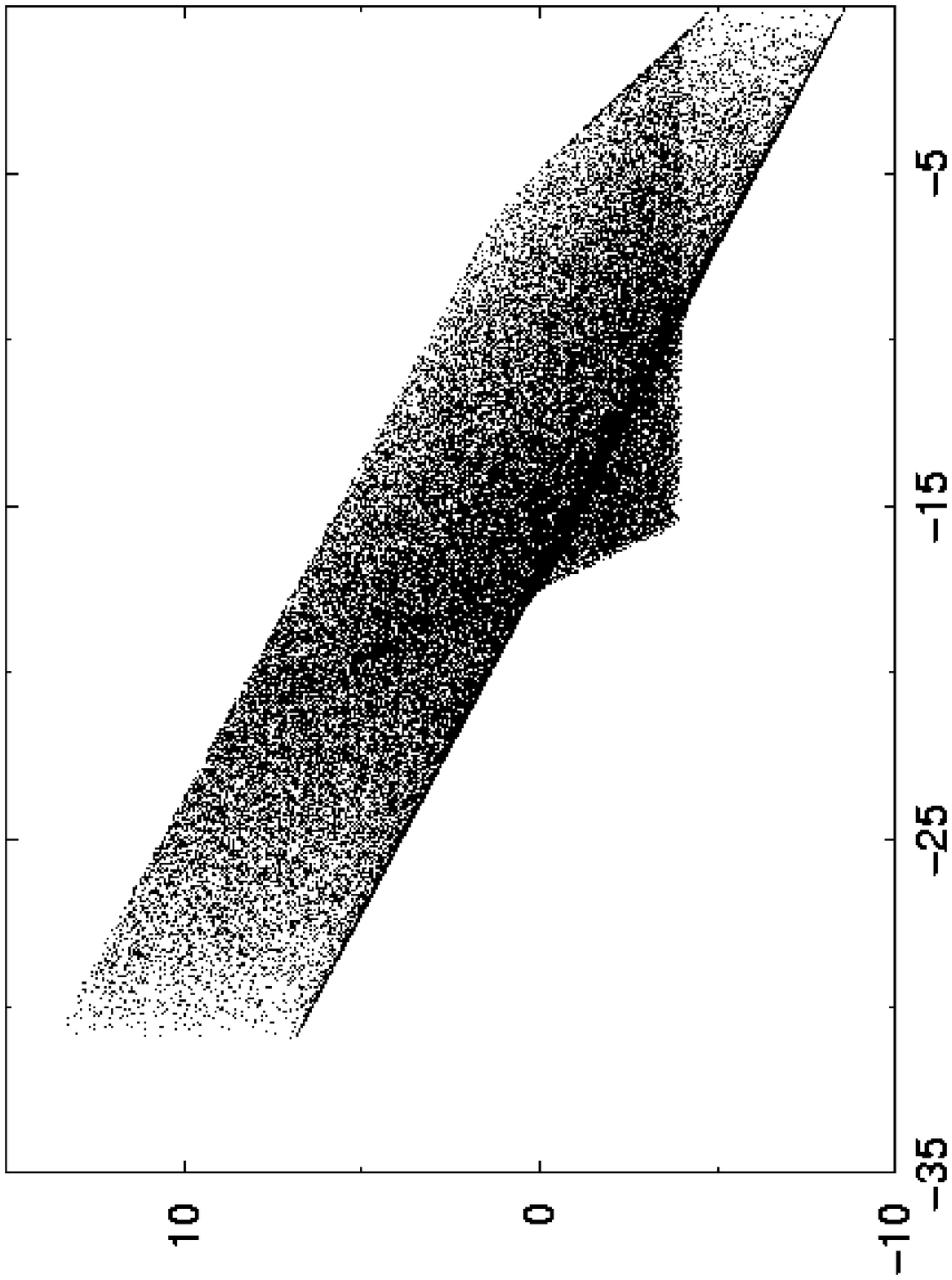}
\put(-110,220){$\scriptstyle \log\frac {\bar R}{\rm mm}$}
\put(-220,325){$\scriptstyle \log\frac {m_2}{\rm eV}$}
  \put(-70,385){(a)}
\includegraphics{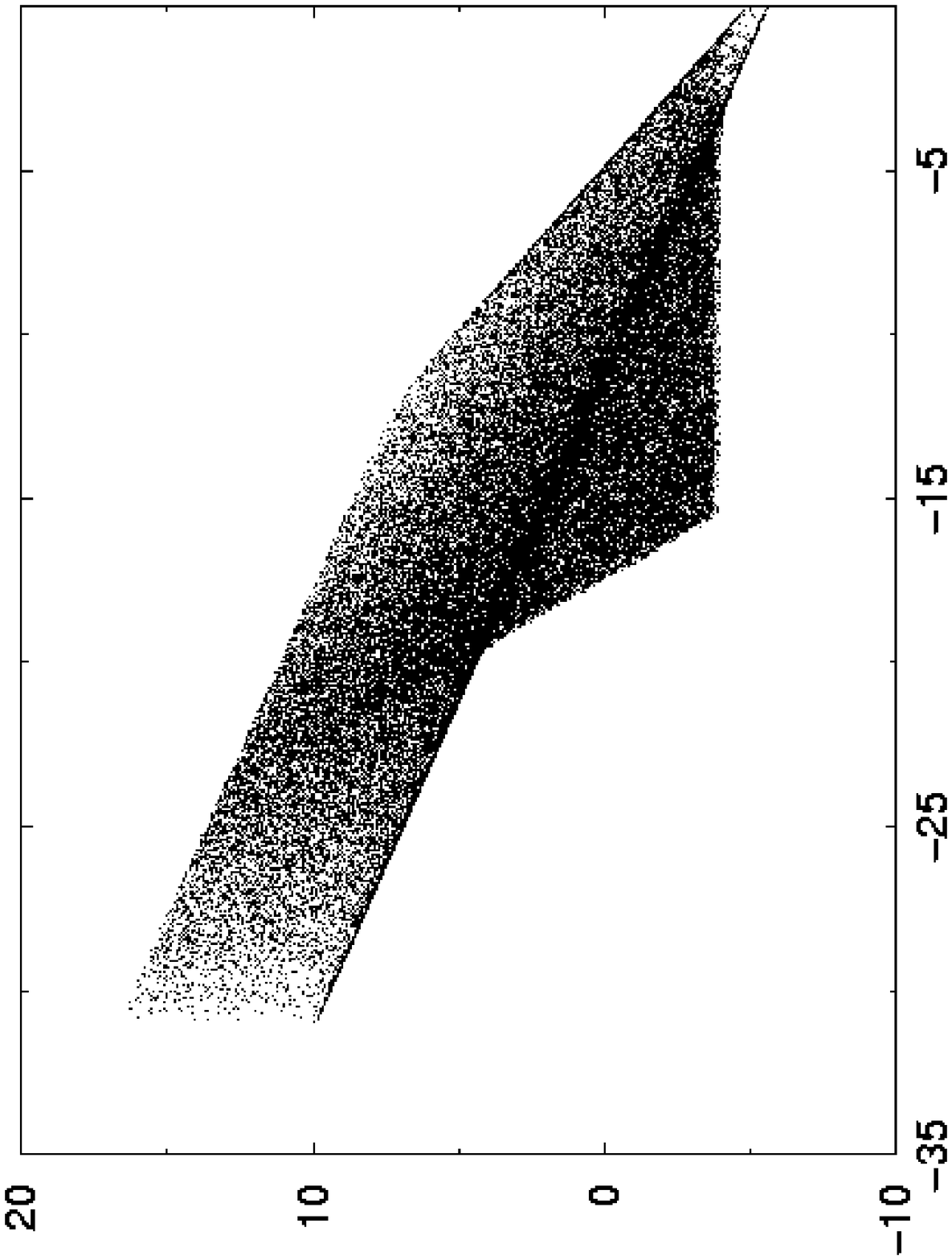}
\put(110,220){$\scriptstyle \log\frac{\bar R}{\rm mm}$}
\put(0,325){$\scriptstyle \log\frac {m_2}{\rm eV}$}
 \put(150,385){(b)}
\end{picture}
\caption{Scattering plot of geometric average of the radii,
$\bar R$, vs.\ possible neutrino masses $m_2$ without
any experimental constraints. a) $\kappa'=0.001$, b)
$\kappa'=1.0$.}
\label{kuva1}
\end{figure}

Figs.\ 3 show how the situation changes when the experimental
constraints are taken into account. Interestingly, the extreme
values on both axes are still allowed and, vaguely speaking, only
dots in between are cut out. Numerical investigations suggest that
the upper (lower) diagonal borderline reflects the chosen range
for the Higgs value $h_3$, higher (lower) values enabling mass
$m_2$ to be still larger (smaller). For the lower diagonal $\sin
\theta \approx 1$. In the case of $\kappa ' =1$ the allowed region
is split up into two separate parts, the feature that will be seen
in all the subsequent figures.

\begin{figure}[ht]
\leavevmode
\centering
\vspace*{100mm}
\begin{picture}(0,0)(0,490)
\includegraphics{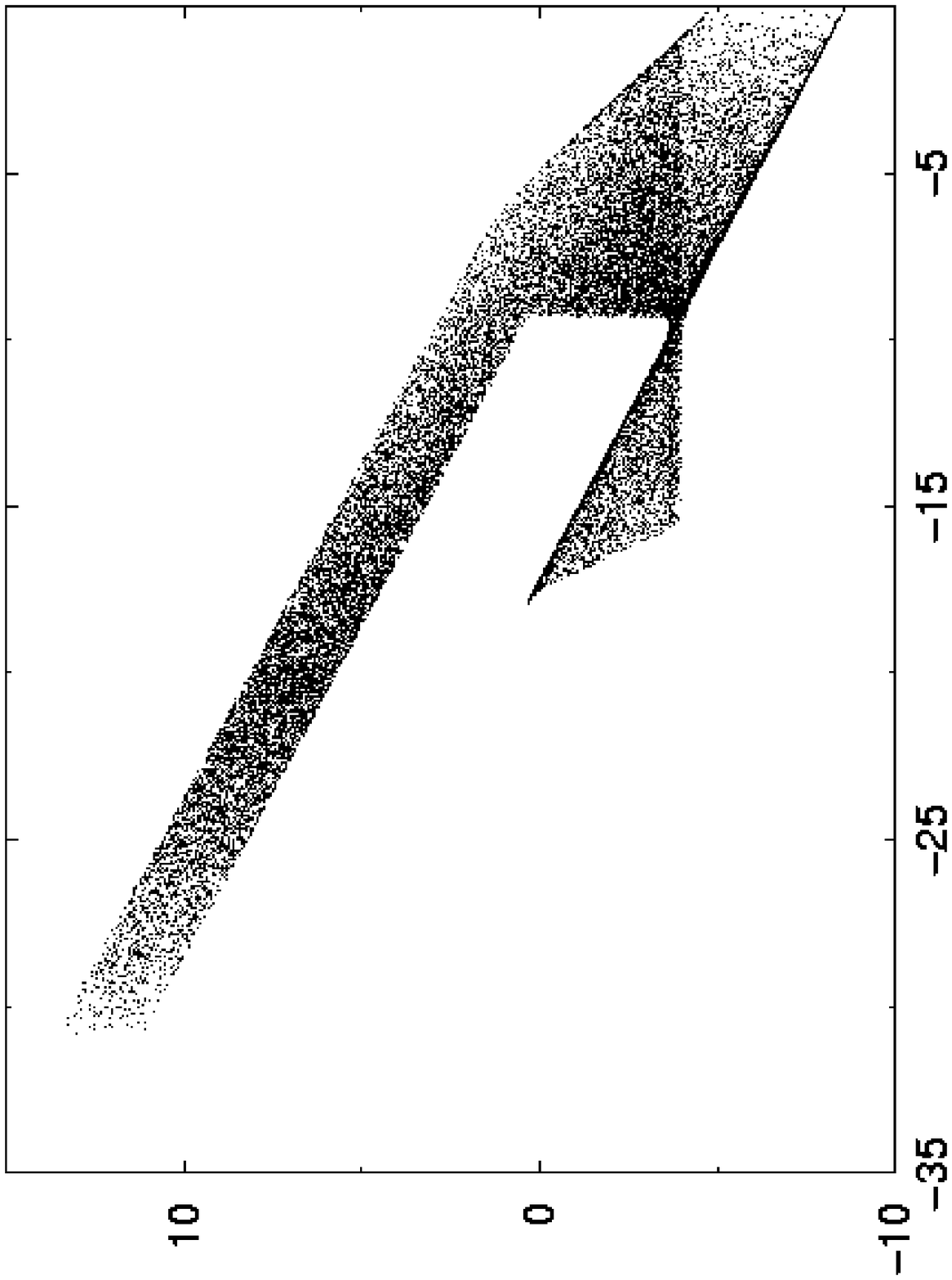}
\put(-110,220){$\scriptstyle \log\frac{\bar R}{\rm mm}$}
\put(-220,325){$\scriptstyle \log\frac {m_2}{\rm eV}$}
  \put(-70,385){(a)}
\includegraphics{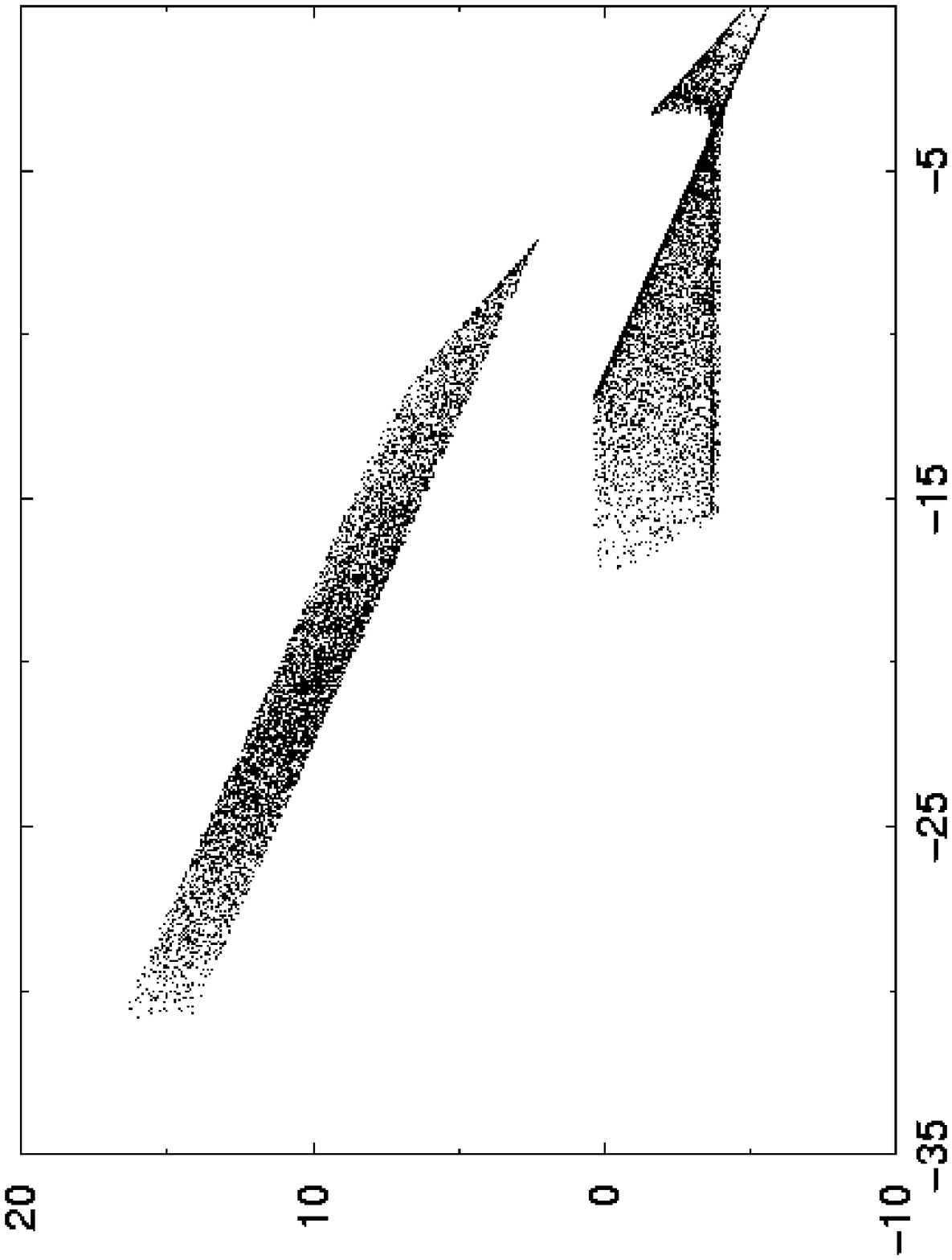}
\put(110,220){$\scriptstyle \log\frac{\bar R}{\rm mm}$}
\put(0,325){$\scriptstyle \log\frac {m_2}{\rm eV}$}
 \put(150,385){(b)}
\end{picture}
\caption{Scattering plot of geometric average of the radii,
$\bar R$, vs.\ possible neutrino masses $m_2$ with
all observational constraints. a) $\kappa'=0.001$, b)
$\kappa'=1.0$.}
\label{kuva2}
\end{figure}

Figs.\ 4 display the behavior of the mixing angle $\theta$ as a
function of $\bar R$. For very small extra-dimensional sizes the
mixing is tiny, but above $\bar R \sim 10^{-16}$ mm maximal
mixing, $\theta =\pi /4$, becomes possible. The peculiarity that
$\theta$ can never be exactly zero is a consequence of the
mathematical structure of the model, as may be deduced from Eq.\
(\ref{tassu}). However, for larger values of $h_3$ the allowed
region would extend to smaller angles. The limit just below $\sin
\theta=10^{-4}$, on the other hand, is due to the cosmological
constraint (\ref{shi}). Note that the obvious change in the
scatter plot for $\bar R \gsim 10^{-10}$ mm ($\kappa' =0.001$) or
$\bar R \gsim 10^{-4}$ mm ($\kappa' =1$), as well as the tilted
rightmost borderline penetrating both of the separate parts in the
case of $\kappa' =1$, have analogous counterparts in Figs.\ 3.

\begin{figure}[ht]
\leavevmode
\centering
\vspace*{100mm}
\begin{picture}(0,0)(0,490) 
\includegraphics{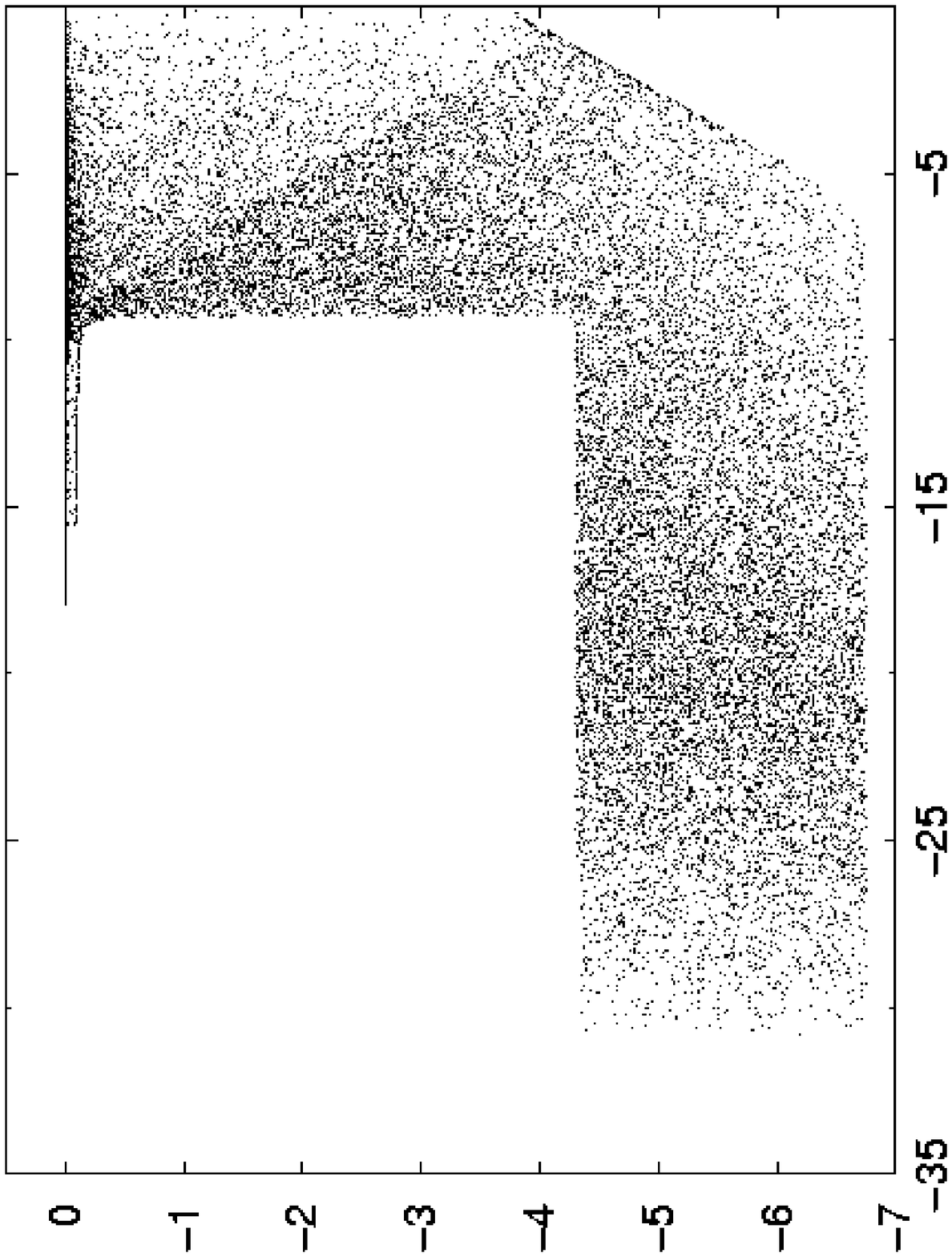}
\put(-110,220){$\scriptstyle \log\frac{\bar R}{\rm mm}$}
\put(-220,325){$\scriptstyle \log\sin\theta$}
  \put(-170,385){(a)}
\includegraphics{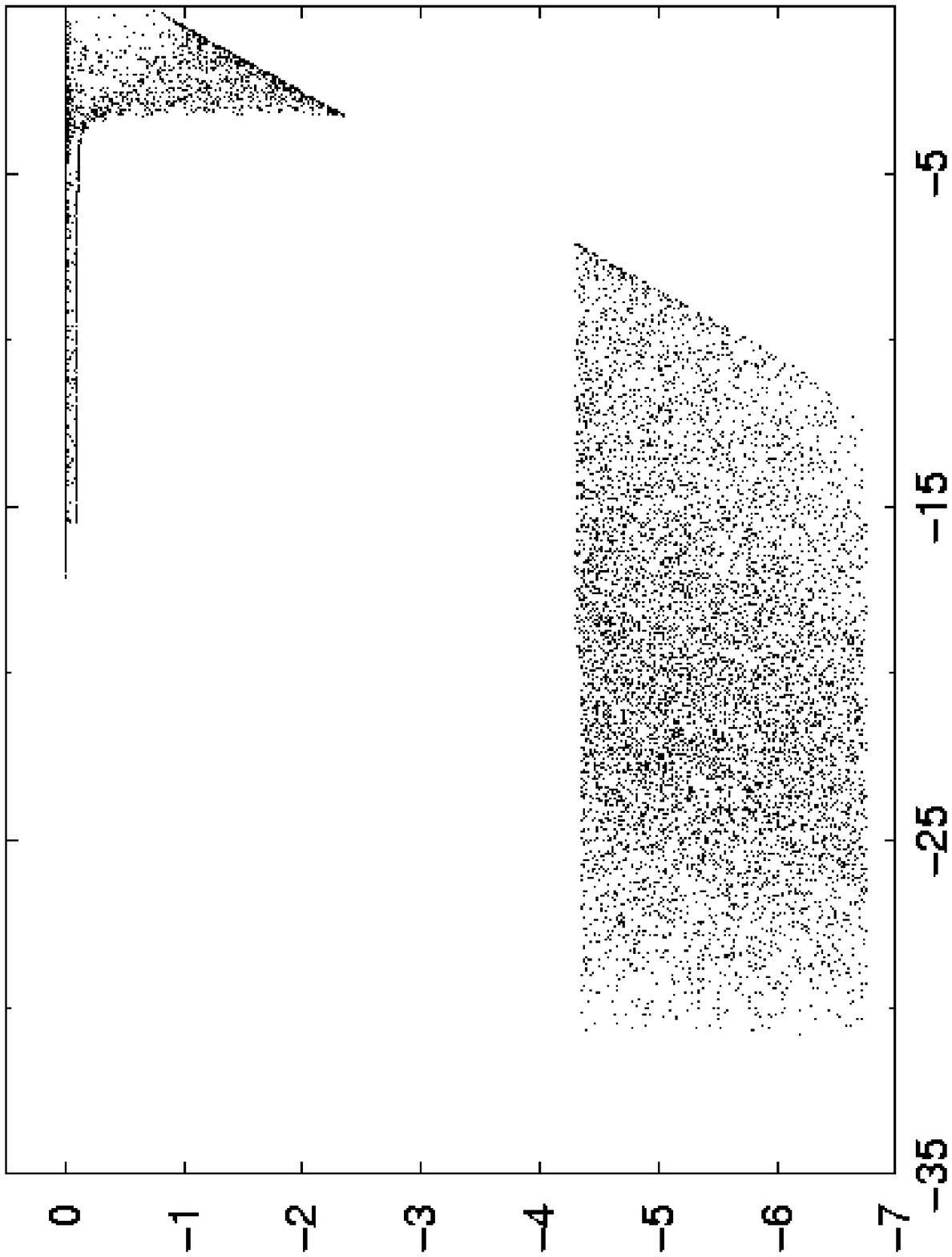}
\put(110,220){$\scriptstyle \log\frac{\bar R}{\rm mm}$}
\put(0,325){$\scriptstyle \log\sin\theta$}
 \put(50,385){(b)}
\end{picture}
\caption{Scattering plot of geometric average of the radii,
$\bar R$, vs.\ mixing angle $\sin\theta$  with
observational constraints. a) $\kappa'=0.001$, b)
$\kappa'=1.0$.}
\label{kuva3}
\end{figure}

Finally, in Figs.\ 5 is shown the allowed parameter space in the
($\sin^{2}2\theta ,\delta m^2$)-plane, where $\delta m^2
=m_{2}^2$ as $m_1 =0$. We have presented in these
figures separately those points which correspond to large values
of the size $\bar R$ of the extra dimension,  0.01 mm $< \bar R <$ 1 mm.
With $\bar R $ in this range, the hierarchy problem can be
solved in this model and also the existence of the extra
dimensions is testable in the new short-range gravity experiments
\cite{4,76}. The empty areas in the large $\sin^{2}2\theta$
large $\delta m^2 $ region are forbidden by the cosmological
constraint (\ref{shi}). The other constraints do not have
substantial effects to this plot. As mentioned before, the
borderline at small $\theta$ (now at $\sin^{2}2\theta\sim
10^{-13}$) corresponds to the upper end of the Higgs vev $h_3$
range covered by our Monte Carlo data set, and it would move
towards smaller values if the
maximum value of the parameter $h_3$ is increased.

As can be seen from Figs.\ 5, our model allows for a vast
diversity in the possible ($\delta m^2 ,\theta$)-combinations. With the
chosen ranges of the free parameters of the model, the squared
mass difference of the active-sterile neutrino pair can have
values from about $10^{-11}$ eV$^2$ ($\kappa'=1$) or $10^{-17}$
eV$^2$ ($\kappa'=0.001$) up to $10^{25}$ eV$^2$ or more. However,
squared mass differences $\delta m^2 =m_{2}^2 \gsim 1$ eV$^2$
are possible only
for very small values of the mixing angle, $\theta\lsim 10^{-4}$.
If the extra dimension size is in the theoretically and
phenomenologically interesting region of $\bar R\gsim 0.01$ mm
(the hits denoted by circles in Figs.\ 5),
$\delta m^2$ is forced to much smaller values.
In particular, maximal
active-sterile mixing is possible in this case only for quite
small values of the squared mass difference: $\delta m^2 \simeq
10^{-11}- 10^{-9} $ eV$^2$ for $\kappa'=1$ and $\delta m^2 \simeq
10^{-17}- 10^{-15} $ eV$^2$ for $\kappa'=0.001$. On the other hand, the
largest squared mass difference consistent with the large extra
dimension size is of the order of $\delta m^2 \simeq 10^{-6} $ eV$^2$.

Let us still mention that the leftmost points in the small squared
mass difference region, $\delta m^2 \lsim 1 \ \mbox{eV}^2$,
correspond to $\theta \approx \pi /2$, as may be seen particularly
well by comparing Figs.\ 4b and 5b. In this case
the predominantly active state, the ordinary
neutrino, is the heavier one of the two mass states.

\begin{figure}[ht]
\leavevmode
\centering
\vspace*{100mm}
\begin{picture}(0,0)(0,490)
\includegraphics{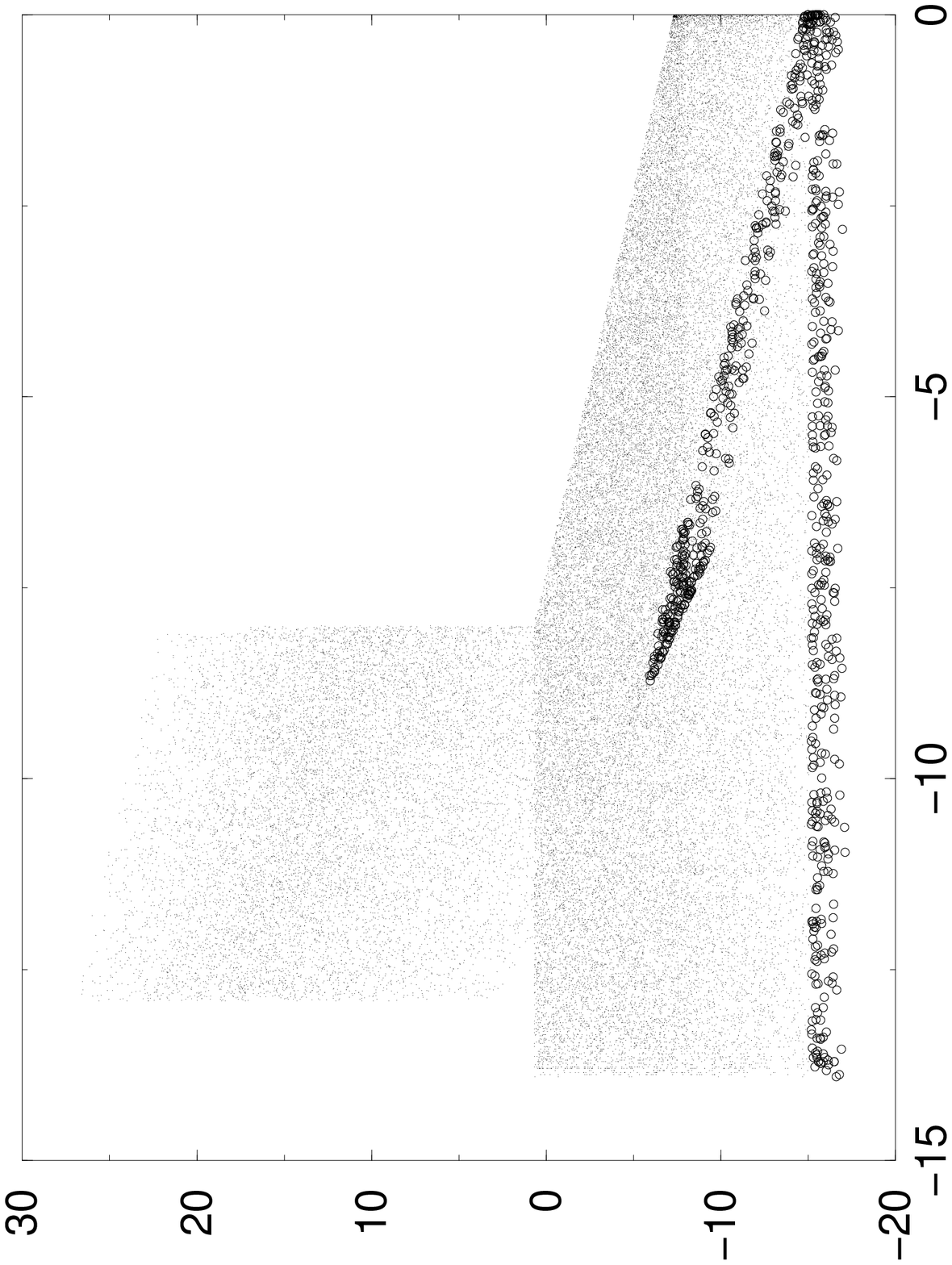}
\put(-110,220){$\scriptstyle \log\sin^2(2\theta)$}
\put(-225,325){$\scriptstyle \log\frac {\delta m^2}{{\rm eV}^2}$}
  \put(-70,385){(a)}
\includegraphics{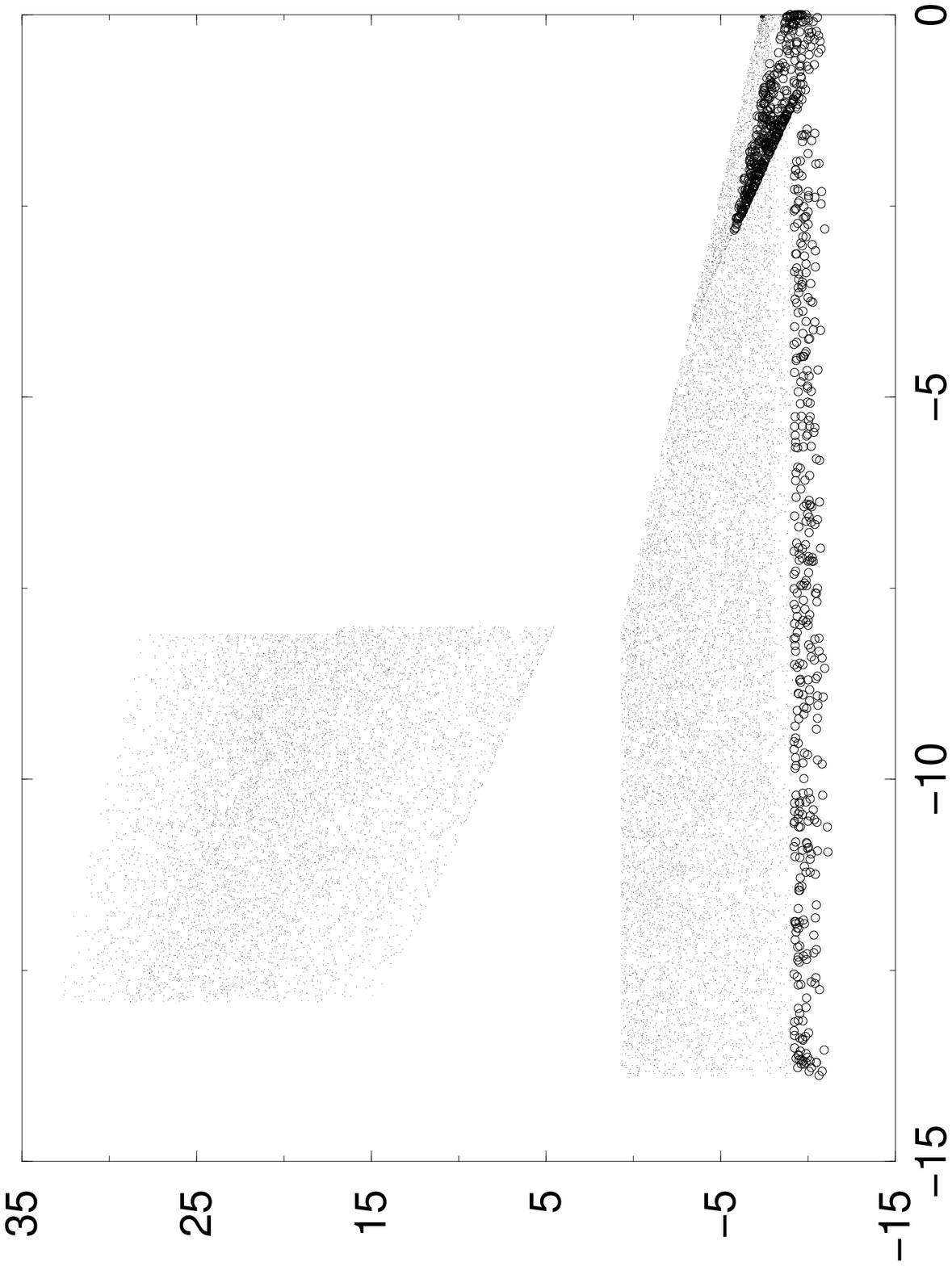}
\put(110,220){$\scriptstyle \log\sin^2(2\theta)$}
\put(-5,325){$\scriptstyle \log\frac {\delta m^2}{{\rm eV}^2}$}
 \put(150,385){(b)}
\end{picture}
\caption{Scattering plot of $\sin^2 (2\theta)$ vs.\
squared neutrino mass difference
$\delta m^2$  with experimental and observational constraints
taken into account. Circles indicate the points, where 0.01 mm $<
\bar R <$ 1 mm. a) $\kappa'=0.001$, b) $\kappa'=1.0$.}
\label{kuva4}
\end{figure}

\section{Summary and conclusions}

We have investigated a six-dimensional model with two extra
dimensions compactified on a torus. Particles with Standard Model
charges are trapped on four four-dimensional branes (vortices) by
a scalar field with an appropriate periodic position dependence on
the torus. On two of the branes, one of which is the world we live
in, particles have the familiar $V-A$ weak currents, while on the
other two, called mirror branes, weak currents are of the $V+A$
form. The structure of the model is mathematically quite
strictly determined, but there are some free parameters in the theory
such as the size of the extra dimensions, characterized by the
radii of the torus $R_1$ and $R_2$, and the vev ($h_3$) of the
counterparts of the SM Higgs field on the other branes. Also the
strength $\kappa'$ of the Yukawa coupling between the Higgs field,
brane neutrinos and bulk neutrinos, as well as the locations of
the vortices on the torus, are not a priori restricted.

Right-handed neutrinos and left-handed mirror neutrinos, which
lack Standard Model interactions, can propagate in the extra
two-dimensional bulk and thereby mediate mixing between active
neutrinos of the branes and mirror branes. This affects
neutrino phenomenology on our brane. We have confronted the model
with existing data on neutrino masses and active-sterile neutrino
mixing from laboratory measurements and cosmology. Our main
results can be seen in Figs.\ 5, where we have plotted the
($\sin^{2}2\theta ,\delta m^2$)-parameter space covered by our model
when the free parameters of the model are allowed to vary within
wide ranges of values. In order the model to solve the hierarchy
problem, the size of the extra dimensions should be large enough.
The corresponding points in Figs.\ 5 are denoted by circles,
and as one can see, the squared mass difference is in
that case forced into a quite narrow range of values, in
particular when the mixing is maximal.

We have restricted our phenomenological analysis to the electron
neutrino, but conclusions would be very similar for the muon and
tau neutrinos, since the most important constraint, the
cosmological bound for the active-sterile mixing,
is quite insensitive to neutrino flavor.

Let us finally emphasize that the model we have considered is not
the most general one but we have made a number of simplifying,
albeit conceivable, assumptions. Nevertheless, we believe that our
analysis has revealed the essential features of the scenario.

\section*{Acknowledgements}

This work has been supported by the Academy of Finland under the
project no.\ 40677. One of us (V.S.) has been funded by the Graduate
School of Nuclear and Particle Physics of the Academy of Finland.

\end{document}